\documentclass[prd,twocolumn,tightenlines,preprintnumbers,showpacs,superscriptaddress,notitlepage,nofootinbib,eqsecnum,floatfix,longbibliography,aps,10pt]{revtex4-1}
\pdfoutput=1
\usepackage[utf8]{inputenc}
\usepackage[T1]{fontenc}
\usepackage{mathrsfs}
\usepackage{bbold}
\usepackage{amsmath}
\usepackage{amssymb}
\usepackage{mathtools}
\usepackage{amsfonts,dsfont}
\usepackage{array}
\usepackage{bm,bbm}
\usepackage{graphicx}
\usepackage{xcolor}
\usepackage{enumitem}
\usepackage{soul}
\usepackage{stmaryrd}
\usepackage{hyperref}
\usepackage{cancel}
\usepackage{tikzsymbols}
\usepackage{diagbox}
\usepackage[normalem]{ulem}
\usepackage{slashed}
\hypersetup{
    colorlinks=true,     
    linkcolor=blue,      
    citecolor=blue,      
    filecolor=blue,      
    urlcolor=blue        
}

\DeclarePairedDelimiter{\ceil}{\lceil}{\rceil}
\DeclarePairedDelimiter{\floor}{\lfloor}{\rfloor} 

\newcommand{\xor}{\oplus}
\newcommand{\eqnref}[1]{Eq.~(\ref{#1})}
\newcommand{\figref}[1]{Fig.~\ref{#1}}
\newcommand{\tabref}[1]{Table~\ref{#1}}

\newcommand{\secref}[1]{Sec.~\ref{#1}}
\newcommand*{\ket}[1]{\left|{#1}\right\rangle}

\newcommand*{\braopket}[3]{\left\langle {#1} \left| {#2} \right| {#3} \right\rangle}

\newcommand*{\swap}[0]{\chi}
\newcommand*{\brgc}{\textrm{brgc}}

\newcommand*{\ferm}{\textrm{ferm}}

\makeatletter
\def\l@subsubsection#1#2{}
\makeatother

\begin{document}
\title{Adiabatic Quantum Computation with the Fermionic Position Space Schr\"odinger Equation}

\author{Kenneth~S.~McElvain}
\email{kenmcelvain@berkeley.edu}
\date{\today}
\affiliation{Department of Physics, University of California, Berkeley, California 94720, USA}
\affiliation{Nuclear Science Division, Lawrence Berkeley National Laboratory, Berkeley, California 94720, USA}
\newcommand{\alert}[1]{\textbf{\color{red}{#1}}}
\renewcommand{\vec}[1]{\boldsymbol{#1}}

\begin{abstract}
The efficient encoding of the fermionic Schr\"odinger equation as a spin system Hamiltonian is a long-term problem.   
I describe an encoding for the fermionic position space Schr\"odinger equation on a finite-volume periodic lattice with a local potential.   
The challenging part of the construction is the implementation of the kinetic energy operator, which is essentially
the Laplacian.   The finite difference implementation on the lattice combines contributions from neighboring lattice sites,
which is complicated by fermionic exchange symmetry.    

Two independently useful techniques developed here are operator filtering and entanglement gadgets.  
Operator filtering is useful when a simple operator acting on a subspace of the full Hilbert space has a desired set of interactions.  Occupation suppression of the complement of the subspace then filters away unwanted contributions of the operator.
Entanglement gadgets encode the same information differently in two sets of qubits.    We may then independently choose the most efficient encoding for operators acting on the qubits.  

The construction for the Laplacian described here has
$\mathcal{O}\left(An 2^D\right)$ cost in bounded Pauli weight terms where $A$ is the number of identical spinless fermions, $N=2^n$ is the number of lattice
points in each direction, and $D$ is the number of dimensions.    The finite volume context protects the gap between the ground state and the first excited state, yielding polynomial time complexity with the box size.

\end{abstract}

\maketitle

\section{INTRODUCTION}
\label{sec:introduction}
The objective of this paper is to formulate the position space Schr\"odinger equation on a lattice for identical fermions with good
scaling as the number of lattice points and the number of particles are increased. 
Good scaling is polynomial in the log of the number of lattice points and linear in the number of particles.   
Such a scaling generates an exponential advantage over classical computing, which is the essential motivation for quantum computing.

While this work focuses on adiabatic quantum computing, Aharonovt et. al. in~\cite{aharonovt2008adiabatic} make the general argument that adiabatic quantum computation is equivalent to gate based computation.    
An example is the transformation of the adiabatic quantum computing formulation of the position space Schr\"odinger equation for distinguishable particles in~\cite{PRXQuantum.3.020356} to a gate based formulation in~\cite{rrapaj2022gate}.

A well-understood way to encode states of identical fermions in second quantization is to associate the occupation of single particle states with qubits, using $\ket{1}$  to indicate occupation.  
A naive set of creation and annihilation operators can be easily constructed $\sigma_j^{+} = (\sigma_j^x + i\sigma_j^y)/2$ and $\sigma_j^{-} =(\sigma_j^x - i\sigma_j^y)/2$.   The states can represent position, momentum, harmonic oscillator, or other states, also carrying spin and isospin, but for this paper, only spinless fermions will be considered.
Extending the naive operators to fermions requires tracking the creation and annihilation phase, which depends on the occupation parity of the states of lessor index.  
Note that the index assignment is arbitrary but must be consistently used.  
Creation and annihilation operators including the required phase factor can be defined as
\begin{equation} \nonumber
a_j^\dagger =({-}1)^j \left[ \prod\limits_{i=0}^{j-1}{ \sigma_{i}^z }\right] \sigma_j^{+} ,\quad a_j =({-}1)^j  \left[ \prod\limits_{i=0}^{j-1}{ \sigma_{i}^z }\right] \sigma_j^{-}
\end{equation}
This transformation of fermionic operators to spin operators is known as the Jordan-Wigner  transform \cite{jordan1928pauli}.
An enhancement with better scaling, logarithmic instead of linear in the number of qubits, is the Bravyi-Kitaev transform~\cite{bravyi2002fermionic}.  It  encodes both occupation and phase information non-locally to simplify creation and annihilation operators, and therefore simplifying Hamiltonians written in terms of them.

One downside of the above encodings for fixed particle number $A$ is the large  overhead associated with representations for  particle numbers other than $A$.    
The movement of a fermion from position index $i$ to position index $j$ can be effected by an operator  $a_j^\dagger a_i$.   
Such an operator carries the complication of tracking the change in the phase of the wave function, which is a function of an assumed ordering of states and the number of occupied single particle states between states $i$ and $j$.    
In one dimension, the position states can be ordered with the indices, resulting in low overhead for hopping from one position to a neighbor.  
However, in higher dimensional problems, a position increment or decrement will result in hopping over many intermediate states, leading to a much higher cost for the kinetic energy operator to account for the phase.

The main goal of this paper is the encoding of a fixed number $A$ of identical fermions in $D$ dimensions with an associated efficient implementation of the Laplacian (essentially the kinetic energy) operator.    
A local potential also benefits because neighboring physical points typically have similar values of the potential.    
In the encoding developed here, the number of qubits and bounded Pauli weight gates will grow polynomially in the log of the number of position states, resulting in an exponential scaling advantage.

Several notational conventions are followed in this paper.  
Tensor products of states and operators are written as simple products.   
Unmentioned qubits in a product are acted on by the identity operator.   
The qubit value 0 corresponds to the spin up state.
The projection operator acting on qubits $a$ and $b$,
\begin{equation}
P_{a,b}^{01} = P_a^0 P_b^1 = \frac{1+ \sigma_a^z}{2} \; \frac{1 - \sigma_b^z}{2} ,
\end{equation}
may be extended to any number of qubits and corresponding binary values.   
With the value $v$ representing a binary number, 
\begin{equation}
P_{q_1,q_2, \ldots, q_n}^v = \prod\limits_{i=1}^n  P_{q_i}^{\floor{v / 2^{i-1}} \textrm{mod}\, 2} .
\end{equation}
Sums of  disjoint projection operators
on the same qubits with distinct binary values are themselves projection operators.    
Projection operators times a weight will be used as penalties to enforce correlations between sets of qubits, in which case sums of
non-disjoint projection operators will generate more efficient and functionally equivalent penalties.

In~\cite{PRXQuantum.3.020356} it was shown how to encode the finite-volume periodic position space distinguishable particle Schr\"odinger equation for adiabatic quantum computation (AQC) in terms of a qubit Hamiltonian so that the representation of $N=2^n$ position states requires $n$ qubits for encoding position in each direction and the Laplacian has a quadratically bounded number of Pauli weight 3 operators.  
An additional benefit described in that reference of working in a finite-volume is that the gap and therefore the adiabatic evolution time is protected by the discrete spectrum of the kinetic energy.

On a lattice with spacing $a$ and discrete positions $a\vec{m}$ generated with integer vector $\vec{m}$, a difference formula approximation to the second derivative is used.
\begin{align}
\nabla^2 \psi(\vec{x}) &\approx  \frac{1}{a^2} \left[ \left(\sum\limits_{m\in \mathcal{N}(n)} \psi(a\vec{m}) \right) -2 D \psi(a\vec{n}) \right] \\
L &= \left(\sum\limits_{m\in \mathcal{N}(n)} \psi(a\vec{m}) \right) ,
\end{align}
where $\mathcal{N}(\vec{n})$  is the set of $\pm 1$ offsets in the $D$ directions, including periodic wrapping in the finite volume.
In the rest of the paper, $L$ will be referred to as the Laplacian for simplicity, but a factor of $-\hbar^2/(2 m a^2)$ will be required when constructing the kinetic energy operator from it.  The central contribution, $2D\psi(a\vec{n})$ will also be omitted as it constitutes a constant diagonal contribution to the Hamiltonian that can be accounted for as a fixed energy shift in the eigenstates.

The encodings used in~\cite{PRXQuantum.3.020356} were the binary reflected Gray code (BRGC) and a Hamming distance 2 Gray code (H2GC).   
With such encodings, neighboring position states have codes differing in exactly one qubit state, simplifying the finite difference implementation of the Laplacian.  
The Laplacian for 4 position states with BRGC codes $00$, $01$, $11$, $10$, ignoring an overall scaling, is simply
\begin{equation}
	L^{(2,\brgc)} = \sigma^x_0 + \sigma^x_1,
\end{equation}
where the 2 in the superscript indicates the number of qubits $n$ used to encode the position with $N=2^n$ positions total.
Extending from $n=2$ to  $n=3$ can be thought of as gluing together two copies with the second in reverse order and correcting the period 4 connections
to period 8.   
Continuing to larger $n$, the corrections are always boundary corrections, isolated by a projection operator.
For the BRGC encoding we have:
\begin{equation} \label{eqn:graylap}
    L^{(n,\brgc)}_{0\ldots n-1} = L^{(n-1,\brgc)}_{0\ldots n-2} +\left(\sigma_{n-1}^x - \sigma_{n-2}^x\right) \prod\limits_{i = 0}^{n - 3} {P_i^0}.
\end{equation}
The Laplacian in this encoding extends linearly to multiple distinguishable particles as a simple sum.   Implementation of
the Laplacian with a BRGC code requires using XZ and ZZ couplings.   
The projection operators in this formulation have a linear cost in $n$, as do the $\sigma^x$ operators.   
The overall BRGC cost in bounded Pauli-weight operators is $\mathcal{O}(A n)$.

The second encoding, H2GC, uses an additional concept to reduce the operator set to X and ZZ operators.   
The cause of XZ couplings in the BRCG encoding is that the  $\sigma^x_i$ operators  used to construct contributions from neighboring position states will also create contributions between distant states.   
These unwanted contributions are suppressed by coupling $\sigma^x_i$ with a projection operator, which generates the XZ coupling use.  
In the H2GC encoding, only a subset of codes represent valid positions states and the unwanted contributions are organized to connect to excluded codes.   
The excluded codes are suppressed by a diagonal penalty contribution to the Hamiltonian which is of quadratically bounded cost.   
The overall H2GC cost in bounded-weight pauli operators is $\mathcal{O}(A n^2 D)$, set by the penalty implementation for excluded basis states.
On the included  basis states, the H2GC Laplacian reduces to a simple sum of $\sigma^x$ operators on the position qubits.

Operator filtering is the generalization of this example.    Suppose we have a set of abstract unencoded states with matrix elements.    Further suppose that an inexpensive operator acting on a qubit product space has a subspace of the same number of states with the same matrix elements.   Then we can filter the operator by suppressing via an occupation penalty the complement of the subspace, thereby ignoring the unwanted  contributions of the operator.  The matching implies the encoding to be used for the abstract  states.  In practice we may find or have an operator that  almost has the desired subspace matrix elements and have to add some corrections to complete it.   This is the situation with the fermionic Laplacian.   A distinguishable particle Laplacian generates a near match with the desired abstract matrix elements with a choice of subspaces.    I choose the subspace and make the required corrections, implying the encoding of the antisymmetric states.  

The fermionic Laplacian to be described below is built on top of a distinguishable particle Laplacian and does not depend on the particular implementation choice except in that the distinguishable particle Laplacian implementation cost, the count of bounded weight Pauli operators, should be $\mathcal{O}(A * poly(\log{N})D)$, to preserve the overall scaling.   
The BRGC and H2GC Laplacians from~\cite{PRXQuantum.3.020356} both satisfy this criterion.   
These Laplacian operators are  invariant under identical particle exchange, so they commute with all exchange operators used in the fermionic construction.  
The Laplacian operators may also be decomposed into independent contributions for each particle.
\begin{equation} \label{eqn:LapParticles}
L = \sum\limits_{i=0}^{A-1} L_i
\end{equation}

The full Hamiltonian to undergo adiabatic evolution, $H=T+V$,  is the sum of the kinetic energy and a local potential.    The
local potential is a diagonal contribution that can be expressed as a sum of products of $\sigma^z$ matrices.  The techniques described
in  ~\cite{PRXQuantum.3.020356} can produce an efficient representation for for the potential, leaving the construction of the
Laplacian as the focus of this work.


\section{Entanglement Gadgets} \label{Sec:Gadget}

Entanglement is often illustrated as a superposition of states of two qubits.
\begin{equation}
\ket{\psi} =\alpha \ket{0}_0 \ket{1}_1 + \beta \ket{1}_0 \ket{0}_1  = \alpha\ket{01} + \beta\ket{10} .
\end{equation}
In this equation, qubits 0 and 1 always have opposite values and information associated with qubit
0 is also associated with qubit 1.    
This entanglement can arise because the Hamiltonian contains a sufficiently large contribution penalizing the other two product states where the qubit values agree.    
The result is that in the low lying states of the overall system will have opposite values for the qubits.

The key idea is that if qubit set $A$ and qubit set $B$ are entangled via  a penalty, then operators acting on common information may be applied to either $A$ or $B$ equivalently.   
The form of the operation on $A$ may differ from the form on $B$ as the information may be represented differently.
This change of representation can yield simpler operations.

A motivating example to generalize from is found in~\cite{PRXQuantum.3.020356}, where the reduction of
a product of two qubit projection operators is implemented by enforcing the entanglement of an ancillary qubit
with a penalty.  
The penalty consists of a weight and a projection operator for the excluded combinations of values.
The table is reproduced in \tabref{tab:reduceprojection}.   
With this penalty in place, $P_a^0$ and $P_{i,j}^{00}$ are equivalent operations on the low-lying states of the system.
A product of such projection operators can therefore be reduced bottom up in a tree with $\ceil{\log_2{n}}$ layers and
$\mathcal{O}(n)$ ancillary qubits.
An alternate top-down approach to subdivision is described in ~\cite{PhysRevA.91.012315, QuantInfComp.8.10.Nov2008},
with the advantage that the left and right-hand operators can be more general.

\begin{table}[h]
\caption{\label{tab:reduceprojection}
A penalty Q entangles qubit a with qubits $i$, $j$.   }
\begin{tabular}{cc|ccc|c}
\hline\hline
\textrm{a} & $P_a^0$ & \textrm{i} & \textrm{j} & $P_{i}^0 P_{j}^0$ & $H_{pen}$ \\
\hline\hline
0 & 1 & 0 & 0 & 1 & 0 \\
0 & 1 & 0 & 1 & 0 & Q \\
0 & 1 & 1 & 0 & 0 & Q \\
0 & 1 & 1 & 1 & 0 & Q \\
1 & 0 & 0 & 0 & 1 & Q \\
1 & 0 & 0 & 1 & 0 & 0 \\
1 & 0 & 1 & 0 & 0 & 0 \\
1 & 0 & 1 & 1 & 0 & 0 \\
\end{tabular}
\end{table}

\subsection{Swap Gadget}
\label{Sec:SwapGadget}
A relevant example for this paper is the action of the swap operator on qubits $i$ and $j$, denoted $\swap_{i,j}$. 
The swap operator is an essential building block for implementing exchange symmetry.
Swap operations have a direct implementation in terms of Pauli operators.
\begin{equation}
	\swap_{i,j} = 
	\begin{bmatrix} 
	   1 & 0 & 0 & 0 \\
	   0 & 0 & 1 & 0 \\
	   0 & 1 & 0 & 0 \\
	   0 & 0 & 0 & 1 \\
	\end{bmatrix}_{i,j}  = \frac{1}{2} \left( I_i I_j + \sigma_i^x \sigma_j^x + \sigma_i^y \sigma_j^y + \sigma_i^z \sigma_j^z \right)
\end{equation}
Native hardware implementations of $\chi$ have begun to appear and even include an extension to a rotation operation on 4 qubits ~\cite{kandel2021adiabatic}.  

In the swap gadget we want to entangle $\ket{c,d}$ with
$\chi_{a,b} \ket{a,b}$ so that actions on the latter can be replaced by actions on the former, simplifying the resulting circuit.
A working penalty contribution with overall strength $Q$ can be written as
\begin{align}
Q (P_{c,a}^{01} &+  P_{c,a}^{10} +P_{d,b}^{01} +  P_{d,b}^{10} )\, \chi_{a,b}  \\
 =& \,Q \big[ \chi_{a,b}- \frac{1}{4} (  \sigma^z_b \sigma^z_c + \sigma^z_b \sigma^z_d +\sigma^z_a \sigma^z_c +\sigma^z_a \sigma^z_d \nonumber \\
 & + i  \sigma^x_a \sigma^y_b \sigma^z_d - i  \sigma^x_a \sigma^y_b \sigma^z_c - i  \sigma^y_a \sigma^x_b \sigma^z_d + i  \sigma^y_a \sigma^x_b \sigma^z_c )\big]. \nonumber 
\end{align}
In the first line the overlapping projectors cover all the cases in which $\ket{c,d}$ does not match $\chi_{a,b} \ket{a,b}$.   
When they match, the penalty contribution is 0.   The overlaps mean that the penalty is a small multiple of $Q$ as
multiple terms may contribute.
A uniform penalty of $Q$ is more expensive to implement.

\subsection{Binary and BRGC}
\label{Sec:Comparison}

A binary comparator is less expensive than a gray code one and the Laplacian is more efficient acting on a BRGC encoding,
so we want both.    The boolean conversion from binary to BRGC can be written in a parallel form, controlling
the depth of the circuit.    In boolean logic we convert from binary code $b_{[n-1:0]}$ to gray code $g_{[n-1:0]}$.
\begin{align*}
g_{n-1} &= b_{n-1}, \\
g_i &= b_i \xor b_{i+1} \;\;(n-1 > i >= 0).
\end{align*}
The correspondence of the gray code qubits below the MSB (most significant bit) is enforced bit by bit by adding a penalty $Q$  for each qubit below index $n-1$ of $g$  to the Hamiltonian when the value of $g_i$ is inconsistent with the values of $b_i$ and $b_{i+1}$.
\begin{equation}
Q \left(P_{g_i,b_i,b_{i+1}}^{100} {+}P_{g_i,b_i,b_{i+1}}^{111}{+}P_{g_i,b_i,b_{i+1}}
^{010}{+}P_{g_i,b_i,b_{i+1}}^{001}\right)
\end{equation}
Each multi-bit coordinate $x_i$ will present in both binary and gray code representations with this mechanism.
The implementation cost for this mechanism is $\mathcal{O}(n A)$.

The point of this  entanglement gadget is that the ordering condition between particles, $x_i < x_{i+1}$, is more easily implemented
on the binary encoding, while the distinguishable particle Laplacian is much more efficiently implemented when the position has a Gray code
representation.   A comparison on the binary representation for two $n$-bit vectors is easily implemented recursively as
\begin{align*}
a_{[n-1:0]} < b_{[n-1:0]} &=  (a_{n-1} \xor b_{n-1}) b_{n-1}\; |  \\
&\quad \overline{ (a_{n-1} \xor b_{n-1})} \left(a_{[n-2:0]} < b_{[n-2:0]}\right), \\
a_0 < b_0 &= P_{a_0,b_0}^{01}
\end{align*}
Each stage in the recursion can be implemented as an ancillary qubit, resulting in a circuit that
has cost $\mathcal{O}(n A)$.    The circuit depth can be reduced from $\mathcal{O}(n)$ to $\mathcal{O}(log(n))$ using standard computer arithmetic techniques
for carry chains~\cite{koren2018computer}, but with higher implementation cost $\mathcal{O}(n \log(n))$.  
In what follows, the operator $a | b$ is the logical or operator, and the operator $a_{[n-1:0]}==b_{[n-1:0]}$ is an  $n$-bit equality comparison operator, which when translated to actions on qubits becomes a projection operator for basis states where the $a$ qubits are in the same states as the $b$ qubits.
Using this, the generalization of the reduced depth comparison 
\begin{align}
a_{[n-1:0]} < b_{[n-1:0]} &=  eq_u \;lt_l \;|\;\overline{eq}_u\; lt_u \\
eq_{u} =& \,a_{\left[n-1:\floor{n/2}\right]} == b_{\left[n-1:\floor{n/2}\right]} \nonumber \\
lt_{u} =& \,a_{\left[n-1:\floor{n/2}\right]} < b_{\left[n-1:\floor{n/2}\right]}\nonumber \\
lt_{l} =& \,a_{[\floor{n/2}-1:0]} < b_{[\floor{n/2}-1:0]},\nonumber 
\end{align}
has a depth proportional to the log of the number of bits being compared.

\section{Identical Fermions}
In the following formulation $A$ fermions exist  in $D$ dimensions with $n$ qubits encoding the axis position for a total of $N=2^{n}$ positions along each axis.
The base qubits are labeled by particle index,  position encoding index and direction : $q_{a,i,d}$, with $a \in [A-1\ldots 0]$, $i \in [n-1 \ldots 0]$, and  $d \in [D-1\ldots 0]$.  We will use $q_a$ to represent the set of bits
\begin{equation} \label{eqn:qabits}
q_a = \left\{q_{a,n-1, D-1}, \ldots, q_{a,n-1,0}, q_{a,n-2, D-1}, \ldots, q_{a,0,0}\right\},
\end{equation}
and $q_{a,d}$ to represent the subset of bits encoding the position of particle $a$ in dimension $d$ in binary.
\begin{equation} \label{eqn:positionbits}
q_{a,d} = \left\{q_{a,n-1, d}, q_{a,n-2, d}, \ldots, q_{a,0,d}\right\}.
\end{equation}
The assignment of 0's and 1's to the bits of $q$ generate the  $2^{ADn}$ basis states of the system.
The left most bit of  \eqnref{eqn:qabits} is considered the most significant bit and the rightmost the least significant in binary comparisons.
This bit assignment is used to order the basis states and to select a representative state for each fermionic Slater determinant.
A key feature of this this bit ordering is that it interleaves the position bits for different directions.   
The advantage of interleaving is that it bounds the the distance (change in position in the ordered basis states) of a increment or
decrement of the position in the D directions, limiting the maximum distance in the ordering to $2^{D-1}$ positions for the $D-1$ direction.  

Additional alternate encoding (i.e. BRGC, or H2GC) qubits suitable for the chosen Laplacian implementation can be produced via
an entangling penalty as shown in \secref{Sec:Comparison}.    
It is also feasible, but more expensive, to directly use BRGC for \eqnref{eqn:positionbits} and still implement the equivalent of binary comparisons. 

For spinless fermions the number of basis states is given by the combination
\begin{equation}
N_{states} = \begin{pmatrix}
2^{n D} \\
A
\end{pmatrix}.
\end{equation}
When $A << nD$ it it much more efficient to track the states of particles than to keep track of state occupation.

\subsection{Fermions in 1 Dimension}
\begin{figure}[ht]
\centering
\includegraphics[scale=0.38 ]{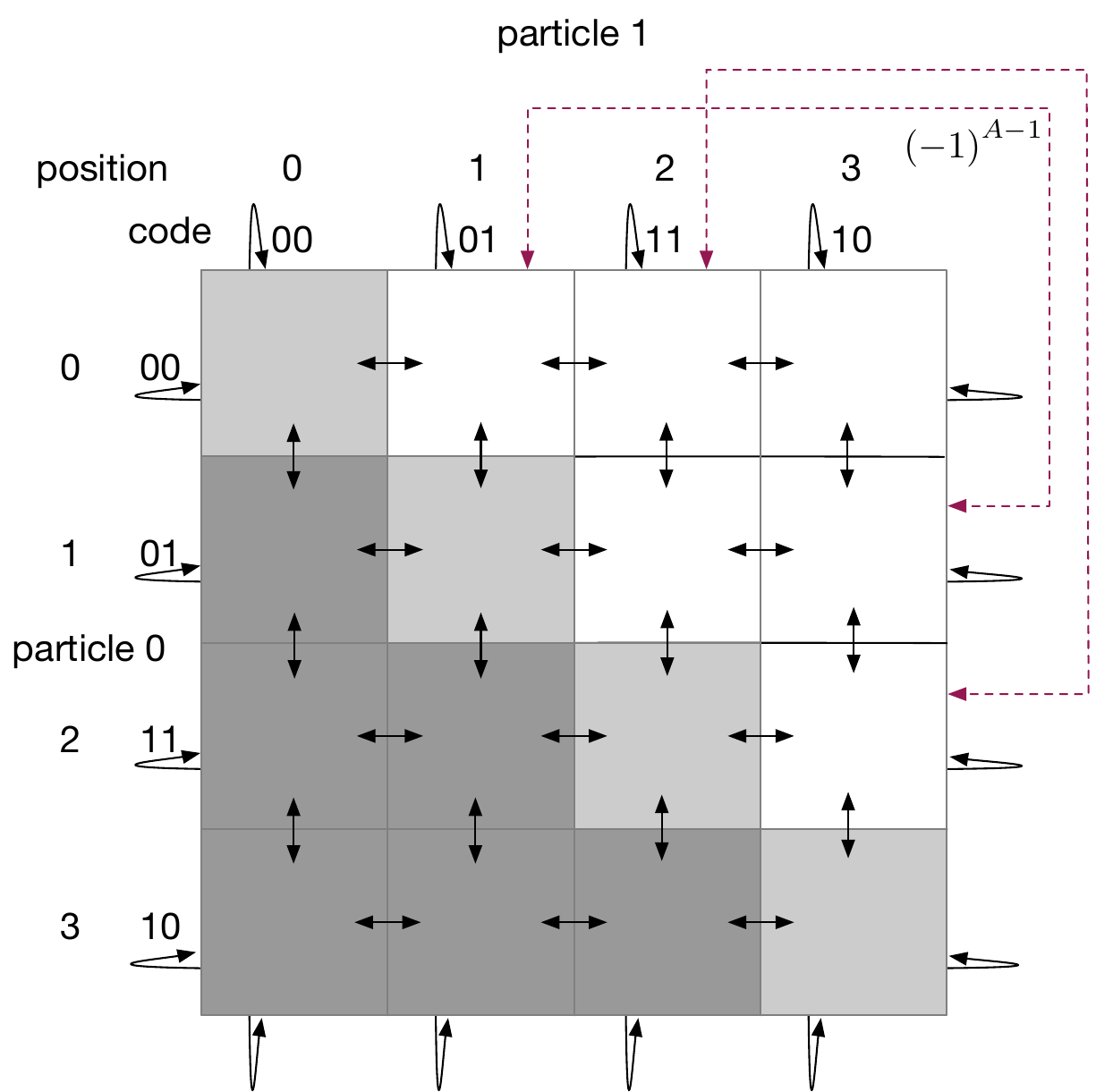}
\caption{ This figure show the implementation of the Laplacian for two identical fermions in 1D  with 4 positions per particle ($n=4$).  Contributions to the Laplacian are shown with solid lines with doubled headed arrows indicating a symmetric contribution between the neighbor states, including periodic wraparound cases.  These are  the contributions for distinguishable particles that are the foundation for the fermionic Laplacian.    One can think of
the amplitude of the square in position $(1,2)$ as representing basis state $\psi_0(1)\psi_1(2)$, or particle 0 in position 1 and particle 1 in position 2, for distinguishable particles.   For fermions the unshaded squares are used to to represent anti-symmetric states and the rest are suppressed with a penalty contribution to the Hamiltonian.  
Boundary corrections completing the connections to neighboring (anti)symmetric states are indicated with dashed lines connecting the top edge to the right edge.   For fermions these connections come with a the phase $(-1)^{A{-}1}$ . }
\label{fig:FermionsNoSpin}
\end{figure}

\figref{fig:FermionsNoSpin} shows how the Laplacian for two fermions in one dimension can be implemented efficiently by a combination of Gray encoding of position,
the use of a penalty contribution to suppress particle label swaps, and the addition of correction terms redirecting the periodic wraparound
contributions to a different edge. 
A penalty contribution suppresses the diagonal squares, which represent identical fermions in the same position state.
A penalty is also applied to the shaded squares below the diagonal, which  represent the particle label exchange partner of a state above the diagonal. 
Unshaded squares  in row $q_0$ and column $q_1$, subject to $q_0 < q_1$, are associated with the anti-symmetric sum 
\begin{equation}
   \frac{1}{\sqrt{2}} \left( \psi_0(q_0) \psi_1(q_1) - \psi_1(q_0) \psi_0(q_1) \right).
\end{equation}
The coordinates for identical fermions states are constrained by
\begin{equation} \label{eqn:coordconstraint}
q_0 < q_1 < \cdots < q_{A-1}
\end{equation}
where the binary bits of $q_i$ are organized according to \eqnref{eqn:qabits}.

In binary or Gray code, the individual comparisons have $\mathcal{O}(Dn)$ implementations, or $\mathcal{O}\left(Dn \log(Dn)\right)$ for implementations with reduced tree height. 
The conjunction of all the comparisons can be reduced to a single ancillary qubit and used to construct a penalty contribution for states violating \eqnref{eqn:coordconstraint}.   
The operation $\textrm{U}$, for unordered,  implements the projector for the violating set of basis states.  
\secref{Sec:Comparison} describes the implementation of the comparisons.
   
A further consequence of this constraint is a simplification of determining the encoding of neighbors.    
Neighbors are found by incrementing or decrementing a single position variable $q_{a,d}$, a part of $q_a$.   
If the initial and final state both satisfy the constraint \eqnref{eqn:coordconstraint}, then that set of $q_a$ is a neighbor.    
If  $A-1 >a > 0$, then the state is not a physical neighbor state (it is a shaded square in the figure) and the Hamiltonian contribution to or from it will be suppressed by the constraint penalty.    
If $q_{0,d}$ is decremented for fermionic state, then there is the possibility that the index wraps and may be  above all other position variables.  
In this case, the coordinates must be rotated to $q_{0}, q_{A-1}, q_{A-2}, \ldots, q_1$ to select the properly ordered neighbor state.   
For three or more particles, a remaining case occurs when $q_{A-1}$ is incremented, in which case the rotation is in the other direction, $q_{A-2}, q_{A-3}, \ldots, q_0, q_{A-1}$.    
The dashed red lines in \figref{fig:FermionsNoSpin} show the neighbor contributions at the edge, which are  associated with the rotation of the coordinate labels.   
The operator $R_{\textrm{wrap}}$ implements the required overall rotations.
\begin{align} \label{eqn:Wrap}
R_{\textrm{wrap}} &= 1 - \chi_{0,1} &\left(A = 2\right) \nonumber \\
 R_{\textrm{wrap}} &= 1 + \left(-1\right)^{A{+}1} \left( R_L +  R_R\right)\quad &\left(A > 2\right)
\end{align}
The phase factors with the number of swaps multiply the rotation operators.    The special case for $A=2$ occurs because $R_L$ and $R_R$ are the same operator on two qubits.
 
Taking advantage of the suppression of the non-physical states in the Hamiltonian the Laplacian operator for 2 or more identical fermions in one dimension can be written
\begin{align} \label{eqn:fbrgc}
L^{\ferm} =& R_{\textrm{wrap}} L
 + Q * \textrm{U} \hfill \nonumber \\
\quad=& L R_{\textrm{wrap}} 
 + Q * \textrm{U},
\end{align}
as $L$ commutes with the rotation operators.  
$Q$ is a penalty weight, and
$R_L$ and $R_R$ are the rotate left and rotate right operators.   
Acting on a properly ordered states, only one of the three operators in $R_{\textrm{wrap}}$ will connect to an unsuppressed state.    

The rotation operators can be implemented as a sequence of $A-1$ pairwise swaps of qubits associated with each particle.  
Since each rotation requires $A-1$ swaps it therefore carries a phase of $(-1)^{A-1}$. 
The rotation operators on $k$ qubits can be implemented as two layers of swap operators.
\begin{equation} \label{eqn:R_R}
R_R =\left( \prod\limits_{i=1}^{\lfloor (k-1)/2 \rfloor} { \swap_{i,k-i} }\right)   
\left( \prod\limits_{i=0}^{\lfloor k/2 \rfloor -1} { \swap_{i,k-i-1} }\right) .
\end{equation}
$R_L$ can be constructed by reversing the bit labeling.    The layers may be replaced by ancillary qubits using the construction in \secref{Sec:SwapGadget} so that
the second layer of swap operators and the application of $L^{\textrm{dist}}$ may be made directly to ancillary qubits.  
 
 \begin{figure}[ht]
\centering
\includegraphics[scale=0.38 ]{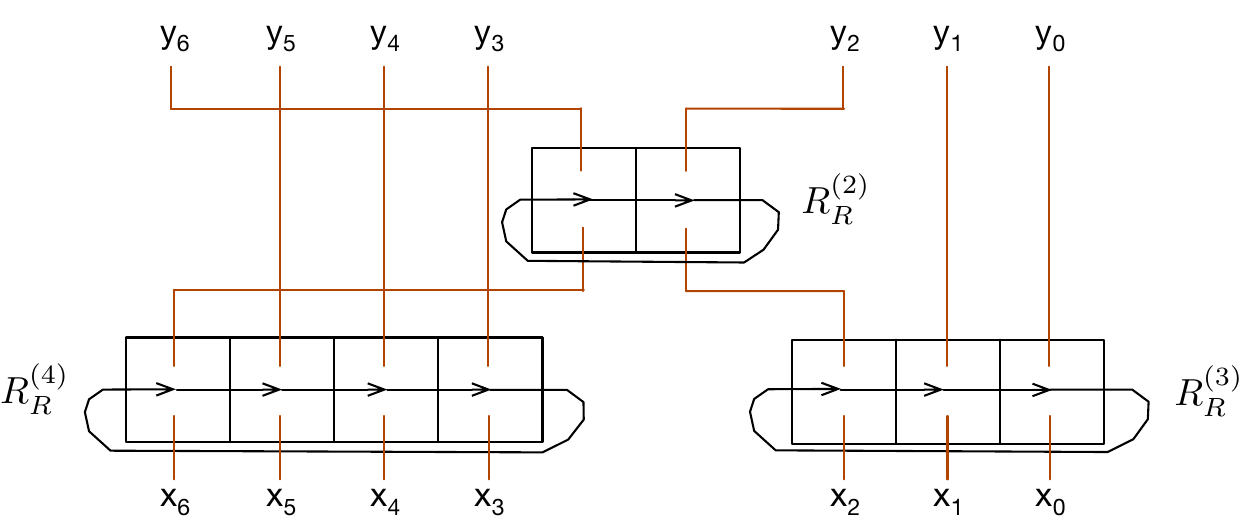}
\caption{ An arbitrary width rotation can be decomposed into a a tree of rotations.   For a right rotation successive layers rotate the leftmost qubits of the previous layer.}
\label{fig:rottree}
\end{figure}
An alternative  implementation for $R_R$ with more layers, but better locality can be constructed recursively by independently rotating a set of partitions followed by a swap of the leftmost qubits of partitions,   see \figref{fig:rottree} for an example.
The base case for the recursion is the trivial one-bit rotation operator, which is simply the identity.  

In later sections, we will require rotations of information, like coordinates, associated with all particles or short sequences of particles.   
$R^R$ and $R^L$ will denote complete rotations of all particles.
The notation $R^R_{3,4,5}$ describes a rotation to the right of particle information for particles 3, 4, and 5.

The distinguishable particle Laplacian $L$ is applied three times: to the original qubits $\left(q\right)$, the ancillary bits for the left rotation $\left(q^{(L)}\right)$,  and the ancillary bits for the right rotation $\left(q^{(R)}\right)$.   The couplings, identified by particle and direction, will connect to an unsuppressed state in at most one of the rotations.

\subsection{Fermions in Multiple Dimensions}

An increment or decrement  of the particle $a$ position in dimension 0 changes $q_a$ by one, finding  either an empty or occupied position.  
If the basis state is unoccupied  and periodic wrapping did not occur, then the new basis state is a valid fermionic state.  
If the position is occupied, then no reordering of the particles will recover a valid fermionic state, yielding no unsuppressed contributions.     
However, incrementing or decrementing positions in dimension d causes a shift in the particle ordering position of $q_a$ of up to $2^d$  due to the interleaving of position bits in \eqnref{eqn:qabits}.   
The result is that a position increment or decrement of  position in dimension d of particle $a$ can hop over up to $2^d-1$ occupied basis states in the basis ordering, resulting in a small group of improperly ordered particles.   When working in two dimensions
 a sum of $A$ pairwise swap operations between adjacent particle indices is sufficient to connect to an ordered fermionic state (up to overall rotations).   
Any swap that does not result in a properly ordered set of particles according to \eqnref{eqn:coordconstraint} will be ignored due to the associated penalty for the basis state.   
The possibility of wrapping also means that the $A$ body rotations of $R_{wrap}$ are still required.
Going forward, particle indices are interpreted as wrapping, mapping $i \rightarrow \left(i\,\textrm{mod}\,A\right)$.
For $A\ge 3$ (to avoid redundant rotation operators), the fermionic Laplacian can be written as
\begin{equation} \label{eqn:TwoDLaplacian}
L^{\ferm}  = L \left(1 {+} \!\sum\limits_{i=0}^{A-1} {  \chi_{i,i{+} 1} }\right) R_{\textrm{wrap}} + Q U .
\end{equation} 
In \eqnref{eqn:TwoDLaplacian} the $L \,1$ represents the case where no particles occupy the states hopped over.  
$L$ is a sum of terms acting independently on each particle.    
If a term acting on a basis state does not result in a hop, then its product with a swap operator can only connect to misordered basis states and is safely ignored.    
If it does cause a hop, then it must be coupled with the swap operator acting on the hopping and hopped particle to contribute.   

The Laplacian $L$ can be decomposed from \eqnref{eqn:LapParticles} into independent pieces over particles labeled with $i$, a direction, and a positive or negative displacement.     
\begin{equation}
L = \sum\limits_{d=0 }^{D-1}   L_{i,d,+} + L_{i,d,-} 
\end{equation}
Then $\eqnref{eqn:TwoDLaplacian}$, in two dimensions, can be rewritten as
\begin{align}
L^{\ferm}  =&\left( L {-} \sum\limits_{i=0}^{A-1} { \left(L_{i,1,+} + L_{i+1,1,-} \right)\chi_{i,i{+} 1} } \right) R_{\textrm{wrap}} \nonumber \\
&+ Q U .
\end{align}
Only displacements in the second dimension (index 1) and above can cause  an improperly ordered state, so only those pieces are combined with the swap
operators.
The local swap operators carry a negative phase.
Using ancillary qubits for all swap and rotation operators results in $\mathcal{O}\left(A n\right)$ additional ancillary qubits.
The ancillary qubits isolate the operators, meaning  their independent costs are additive.
The overall cost in bounded weight terms is then $\mathcal{O}\left(A  n\right)$.

In more dimensions, the increment of dimension $d=2$ position moves four slots in the basis order, and can therefore hop over up to three possibly occupied basis states, requiring a four-position rotation to recover proper ordering.    
In D dimensions, the  increment of the last dimension position moves $2^{D-1}$ slots, requiring all local rotations of up to $2^{D-1}$ particles in both directions.    
For $D = 3$ and $A >= 5$, 
\begin{align}  \label{eqn:3DLaplacian}
L^{\ferm}  &=Q U + \big[ L \nonumber \\
 -& \sum\limits_{i=0}^{A-1} { \sum\limits_{d=1}^2 \left(L_{i,d,+} +L_{i+1,d,-} \right)\chi_{i,i{+} 1} } \nonumber \\
 +& \sum\limits_{i=0}^{A-1} { L_{i,2,+} \; R_{i,\ldots,i+2}^{R}+L_{i+2,2,-} \; R_{i,\ldots,i+2}^{L} } \nonumber \\
  -& \sum\limits_{i=0}^{A-1} { L_{i,2,+} \; R_{i,\ldots,i+3}^{R} +L_{i+3,2,-} \; R_{i,\ldots,i+3}^{L}} \nonumber \\
&\big]  R_{\textrm{wrap}}  .
\end{align}
For $A<5$ redundant rotation operators must be removed, just as in \eqnref{eqn:Wrap} where for $A=2$ the $R_L$ and $R_R$ operators become redundant.
Illustrative examples are the $A=2$ rotation by two or four positions, which are equivalent to a rotation by zero positions.   There must be exactly one way to connect an improperly ordered state to a properly ordered one.

The overall cost becomes $\mathcal{O}\left(A  n 2^D\right)$.
For more dimensions, one  extends the sizes of the rotation operators, coupling them with the pieces of the Laplacian that can cause matching displacement sizes.
The cost grows exponentially with D, but D is generally a small fixed number.

\section{The Potential}
With multiple species of fermions, such as neutrons and protons, where the particle number of each species is fixed, the potential is easily implemented.
For example we could have species $n$ and $p$ representing neutrons and protons with $A_n$ and $A_p$ being the respective particle counts for the species.
System states will be a tensor product of Slater determinants for each species.
Each included state is a sum over all particle permutations, so the neutron-neutron contribution would be
\begin{equation}
\braopket{\psi_a}{\sum_{i<j \in \textrm{species}_n} \!\!\!\!\!V_{i,j}}{\psi_b} = \frac{A_n(A_n{-}1)}{2} \braopket{\psi_a}{V_{n_0,n_1}}{\psi_b},
\end{equation}
as only the interaction between the first two neutrons is needed.

Between species the result is similar
\begin{equation}
\braopket{\psi_a}{\sum_{\substack{i \in \textrm{species}_n\\j \in \textrm{species}_p}} V_{i,j}}{\psi_b} = A_n A_p \braopket{\psi_a}{V_{n_0,p_0}}{\psi_b}
\end{equation}

\section{Summary and Conclusion}
\label{sec:summary}
In~\cite{PRXQuantum.3.020356}, the idea of embedding the basis states of a system in a larger Hilbert space with occupation suppression of the
extra states was used to simplify the operator structure of the position space Laplacian for distinguishable particles.   
Here a generalization of the concept has a new use, embedding the ordered basis states of an A-body identical fermion system in a Hilbert space supporting all particle
orderings, but with a penalty suppressing occupation of improperly ordered states.  
The specific proper particle ordering used here is carefully chosen so that local motions of particles associated with the Laplacian operator have a bounded associated distance in the basis state ordering.   
The bounded ordering distance means that bounded local rotations plus overall rotations for the wrapping case are sufficient to restore the proper ordering of particles after local motion in the lattice.  The improperly ordered and suppressed basis states in the oversized Hilbert space act as trash can, discarding all but one of the local permutation network results and greatly simplifying the required circuit.

Entanglement gadgets are also introduced,  requiring two sets of qubits to carry common information with different encodings.
This enables matching the encoding of information to local requirements in the circuit.   
For example, in the fermionic Laplacian, the proper ordering projector is more efficient when positions are encoded in a binary format, while the distinguishable particle Laplacian used in the construction is dramatically more efficient operating on positions encoded in Gray code.   
Entanglement gadgets let us have both, and are expected to have broader application.

Support for bosons is a non-trivial extension because a position increment can potentially hop over an arbitrary number of particles, removing
the bound on the size of the local rotation operators.
Some important future work is the extension to internal degrees of freedom such as spin and isospin.    

The use of these techniques bounds the cost of the dominating permutation network to $O(A n 2^{D-1})$, yielding an exponential space advantage over classical computing.
Assuming a local potential and initial state preparation for the free fermion wave function, the  potential can be introduced at a rate protected by the finite volume kinetic energy gap to the first excited state and one can expect the evolution time to scale polynomially with the box size.


\section{ACKNOWLEDGEMENTS}

Lawrence Berkeley National Laboratory (LBNL) is operated by The Regents of
the University of California (UC) for the U.S. Department of Energy (DOE) under
Federal Prime Agreement DE-AC02-05CH11231.
This material is based upon work supported by the U.S. Department of Energy,
Office of Science, Office of Nuclear Physics,
Quantum Horizons: QIS Research and Innovation for Nuclear Science
under Award Number FWP-NQISCCAWL.

\bibliographystyle{apsrev4-1}
\bibliography{fermions.bib}
\end{document}